\documentclass{PoS}
\usepackage[inline]{enumitem}
\usepackage{etex}
\usepackage{comment}
\usepackage{amsthm}
\usepackage{dcolumn}
\usepackage{epsf}
\usepackage{mathrsfs}
\usepackage{multirow}
\usepackage{booktabs}
\usepackage{tabularx}
\usepackage{array}
\usepackage{slashed}
\usepackage{float}
\usepackage{leftidx}
\usepackage{setspace}
\usepackage{verbatim}
\usepackage{adjustbox}
\usepackage{tikz}
\usepackage[caption=false]{subfig}
\usepackage{arydshln}
\usepackage{array}
\usepackage{amsmath}

\newcommand{\bea}{\begin{eqnarray}}
\newcommand{\eea}{\end{eqnarray}}
\newcommand{\bean}{\begin{eqnarray*}}
\newcommand{\eean}{\end{eqnarray*}}

\def\W #1{\widetilde{#1}}

\def\Label#1{\label{#1}}

\newcommand{\ctobedelete}[1]{}
\DeclareMathOperator{\rank}{rank}
\title{Correspondence between Solutions of Scattering Equations and Scattering Amplitudes in Four Dimensions}

\ShortTitle{Solutions of SE and Scattering Amplitudes in 4d}

\author{\speaker{Yi-Jian Du}
\\
        Center for Theoretical Physics,
School of Physics and Technology,
Wuhan University,\\
299 Bayi Road, Wuhan 430072,
China\\
        E-mail: \email{yijian.du@whu.edu.cn}}

\author{Fei Teng\\
       Department of Physics and Astronomy, University of Utah,\\ 115 South 1400 East, Salt Lake City, UT 84112, USA\\
     E-mail: \email{Fei.Teng@utah.edu}}

\author{Yong-Shi Wu \\Department of Physics and
Center for Field Theory and Particle Physics, Fudan University,\\
220 Handan Road, Shanghai 200433, China\\
      Department of Physics and Astronomy, University of Utah,\\ 115 South 1400 East, Salt Lake City, UT 84112, USA\\
     E-mail: \email{wu@physics.utah.edu}}

\abstract{In this talk, we review our recent work on direct evaluation of tree-level MHV amplitudes by Cachazo-He-Yuan (CHY) formula. We also investigate the correspondence between solutions to scattering equations and amplitudes in four dimensions along this line. By substituting the MHV solution of scattering equations  into the integrated CHY formula, we explicitly calculate the tree-level MHV amplitudes for four dimensional Yang-Mills theory and gravity. These results naturally reproduce the Parke-Taylor and Hodges formulas. In addition, we derive a new compact formula for tree-level single-trace MHV amplitudes in Einstein-Yang-Mills theory, which is equivalent to the known Selivanov-Bern-De Freitas-Wong (SBDW) formula.
Other solutions do not contribute to the MHV amplitudes in Yang-Mills theory, gravity and Einstein-Yang-Mills theory. We further investigate the correspondence between solutions of scattering equation and helicity configurations beyond MHV and proposed a method for characterizing solutions of scattering equations. }

\FullConference{38th International Conference on High Energy Physics\\
		3-10 August 2016\\
		Chicago, USA}

\begin{document}

\section{Introduction}
The Cachazo-He-Yuan (CHY) formula \cite{Cachazo:2013gna,Cachazo:2013hca,Cachazo:2013iea, CHY-EYM} is a highly compact one for scattering amplitudes.
Compared to Feynman diagrams, the CHY formula may reveal more hidden simplicities of quantum field theory. It has been proposed to exist in many theories including Yang-Mills theory, gravity and Einstein-Yang-Mills theory. The formalism states that $n$-point tree-level massless amplitudes $A_n$ in arbitrary dimensions can be expressed as
\begin{equation}
	A_{n}=\sum\limits_{\{\omega\}\in\,\text{sol.}}\frac{\mathcal{I}_{n}(\{k, \epsilon, \omega\})}{{\det}'[\Phi(\{k,\omega\})]}\,, \Label{eq:CHY}
\end{equation}
where the sum is over all possible $(n-3)!$ solutions $\{\omega\}$ of the scattering equations
\begin{align}
	&\sum_{\substack{b=1\\b\neq a}}^{n}\frac{s_{ab}}{z_{ab}}=0\,,\qquad a\in\{1,2,\ldots,n\}\,,
\Label{eq:SE}
\end{align}
which are M\"obius covariant \cite{Cachazo:2013gna,Cachazo:2013hca,Cachazo:2013iea}. Here $z_{ab}\equiv z_a-z_b$ while  $s_{ab}\equiv2k_a\cdot k_b$ are the Mandelstam variables. In the CHY formula \eqref{eq:CHY},  external polarizations $\epsilon$ are packaged into the CHY integrand $\mathcal{I}_{n}(\{k, \epsilon, \omega\})$ which relies on theories. The co-rank $3$ matrix $\Phi_{ab}$ in \eqref{eq:CHY} is defined by
\begin{equation}
  \Phi_{ab}=\frac{s_{ab}}{\omega_{ab}^{2}}\quad(a\neq b),\qquad\Phi_{aa}=-\sum_{c\neq a}\frac{s_{ac}}{\omega_{ac}^{2}}\,.
  \Label{eq:Phi}
\end{equation}
Deleting arbitrary three rows $(i,j,k)$ and columns $(p,q,r)$ from $\Phi_{ab}$, we get a $(n-3)\times (n-3)$ submatrix $\Phi_{pqr}^{ijk}$. The reduced
determinant ${\det}'(\Phi)$ in \eqref{eq:CHY} is defined by
\begin{equation}
   {\det}'(\Phi)\equiv(-1)^{i+j+k+p+q+r}\frac{\det\left(\Phi_{pqr}^{ijk}\right)}{\omega_{ij}\omega_{jk}\omega_{ki}\omega_{pq}\omega_{qr}\omega_{rp}}\,,\qquad(i<j<k\text{ and }p<q<r)\,.\Label{eq:det}
\end{equation}
Actually, ${\det}'(\Phi)$ is independent of the choice of $(i,j,k)$ and $(p,q,r)$~\cite{Cachazo:2013hca}. The solutions of scattering equation \eqref{eq:SE} play as a critical role in understanding the integrated CHY formula \eqref{eq:CHY}. In four dimensions, there are two special solutions \cite{Solutions,Solutions2}
\begin{equation}
\sigma_{a}=\frac{\langle a\eta\rangle\langle\theta\xi\rangle}{\langle a\xi\rangle\langle\theta\eta\rangle}\,,\qquad\overline{\sigma}_{a}=\frac{[a\eta][\theta\xi]}{[a\xi][\theta\eta]}\,,
\end{equation}
where the arbitrary projective spinors $\eta$, $\theta$ and $\xi$ encode the M\"obius freedom in the solutions. It was conjectured \cite{Solutions2} that only the  special solution $\{\sigma_a\}$\footnote{This solution is also mentioned as MHV solution.} (or $\{\overline{\sigma}_a\}$) contributes to the maximally-helicity-violating (MHV) (or $\overline{\text{MHV}}$) amplitudes in Yang-Mills theory and gravity.

In this talk, we review direct evaluation of tree amplitudes using the CHY formula  \eqref{eq:CHY} in four dimensions.
By plugging the MHV solution $\{\sigma_a\}$ into the CHY formula, we derive the well known Parke-Taylor \cite{Parke:1986gb} and Hodges formulas \cite{Hodges:2012ym} in Yang-Mills theory and gravity. We also derive a new compact formula for tree-level single-trace MHV amplitudes which is equivalent to Selivanov-Bern-De Freitas-Wong (SBDW) formula \cite{SBDW}. Thus the correspondence between the special solution $\{\sigma_a\}$ and MHV amplitudes in Yang-Mills theory, gravity and Einstein-Yang-Mills (EYM) theory are explicitly proved. The general correspondence between solutions to scattering equations and amplitudes beyond MHV in four dimensions are further discussed.

\section{Direct evaluation of MHV amplitudes in Yang-Mills theory and gravity by CHY}
To establish the correspondence between the special solution $\{\sigma_a\}$ and MHV amplitudes, we start from the following CHY integrands for Yang-Mills theory and gravity
\begin{align}
	&\mathcal{I}_{n}(\{k,\epsilon,\omega\})=\frac{\text{Pf}\,'[\Psi(\{k,\epsilon,\omega\})]}{\omega_{12}\omega_{23}\ldots \omega_{n1}}& &\text{color-ordered Yang-Mills amplitudes,}\nonumber\\*
	&\mathcal{I}_{n}\left(\{k,\epsilon,\W{\epsilon},\omega\}\right)={\text{Pf}\,'\left[\Psi(\{k,\epsilon,\omega\})\right]\times\text{Pf}\,'\left[\Psi(\{k,\W{\epsilon},\omega\})\right]}& &\text{gravity amplitudes,}
\Label{eq:YMgravity}
\end{align}
where we use  $\{k\}$ to denote the set of external momenta and $\{\epsilon\}$ (both $\{\epsilon\}$ and $\{\W{\epsilon}\}$) the set of polarizations of external gluons (gravitons). The $2n\times 2n$ antisymmetric matrix $\Psi$ and the reduced Pfaffian are defined by
\begin{equation}
	\Psi(\{k,\epsilon,\omega\})=\left(\begin{array}{cc}
		A & -C^{T} \\
		C & B \\
		\end{array}\right)\,,\qquad\text{Pf}\,'(\Psi)=\frac{(-1)^{i+j}}{\omega_{ij}}\text{Pf}\,(\Psi_{ {i} {j}}^{ {i} {j}})\,,\quad(1\leqslant i<j\leqslant n)\,,
\Label{eq:Psi}
\end{equation}
where the blocks are
\begin{align}
& A_{ab}=\left\{\begin{array}{>{\displaystyle}c @{\hspace{1em}} >{\displaystyle}l}
\frac{s_{ab}}{\omega_{ab}} & a\neq b\\
0 & a=b \\
\end{array}\right.\,,&
& B_{ab}=\left\{\begin{array}{>{\displaystyle}c @{\hspace{1em}} >{\displaystyle}l}
\frac{2\epsilon_{a}\cdot\epsilon_{b}}{\omega_{ab}} & a\neq b\\
0 & a=b \\
\end{array}\right.\,,&
& C_{ab}=\left\{\begin{array}{>{\displaystyle}l @{\hspace{1em}} >{\displaystyle}l}
\frac{2\epsilon_{a}\cdot k_{b}}{\omega_{ab}} & a\neq b\\
-\sum_{c\neq a}\frac{2\epsilon_{a}\cdot k_{c}}{\omega_{ac}} & a=b \\
\end{array}\right.\,.
\Label{eq:ABC}
\end{align}
The upper half part of $\Psi$, $(A,-C^{T})$, has two null vectors such that we need to delete two rows and columns in the first $n$ rows and columns to obtain a nonzero Pfaffian. The reduced Pfaffian is independent of the choice of $(i,j)$ and is permutation invariant.
The Parke-Taylor like factor $(\omega_{12}\omega_{23}\ldots \omega_{n1})^{-1}$ in \eqref{eq:YMgravity} encodes permutation of gluons.

For the MHV  Yang-Mills amplitude with the two negative helicity gluons at $x$ and $y$, we choose the gauge of external polarizations as follows
\begin{align}
    \epsilon_{i}^{\mu}(-)=\frac{\langle i|\gamma^{\mu}|n]}{\sqrt{2}[ni]}\quad(i=x,y)\,,&
    &\epsilon_{j}^{\mu}(+)=\frac{\langle x|\gamma^{\mu}|j]}{\sqrt{2}\langle xj\rangle}\quad(1\leq j\neq x,y\leq n)\,,\Label{eq:gauge}
\end{align}
for convenience. The reduced Pfaffians are independent of the choice of gauge \cite{Cachazo:2013gna,Cachazo:2013hca,Cachazo:2013iea}. For gravity amplitudes, we choose $\epsilon_i^{\mu\nu}(\pm)\to \epsilon_i^{\mu}(\pm)\epsilon_i^{\nu}(\pm)$. Now we substitute the special solution $\{\sigma_a\}$ into the reduced Pfaffian \eqref{eq:Psi}, the reduced determinant \eqref{eq:det} and the Parke-Taylor like factor. Using elementary transformations, we prove the following identities  \cite{Du:2016blz}
\begin{subequations}
\begin{align}
        &\textrm{Pf}\,'(\Psi) =[F(\xi,\eta,\theta)]^{n}(P_{\xi})^{2}{\langle xy\rangle^{4}}{\bar{M}(12\ldots n)},& &F(\xi,\eta,\theta)\equiv\frac{\langle\theta\eta\rangle}{\langle\eta\xi\rangle\langle\theta\xi\rangle}\,,\\
  &{\det}'(\Phi) =[F(\xi,\eta,\theta)]^{2n}(P_{\xi})^{4}{\bar{M}(12\ldots n)},& &P_{\xi}\equiv\prod_{a=1}^{n}\langle a\xi\rangle\,,
\end{align}
\begin{align}
       &\sigma_{12}\cdots\sigma_{n1} =\left[{F(\xi,\eta,\theta)}\right]^{-n}(\langle 12\rangle\cdots\langle n1\rangle)/(P_{\xi})^{2}\,.
\end{align}
\label{eq:Identities}
\end{subequations}
%
The $\bar{M}(12\ldots n)$ is Hodges' reduced amplitudes (see \cite{Hodges:2012ym}), which is gauge invariance and defined by
\bea
			\bar{M}(12\ldots n)&=&(-1)^{n+1}\frac{(-1)^{i+j+k+p+q+r}}{\langle ij\rangle\langle jk\rangle\langle ki\rangle\langle pq\rangle\langle qr\rangle\langle rp\rangle}\det(\phi_{pqr}^{ijk}),\nonumber
			\eea
where the definition of $\phi_{ab}$ matrix is
\begin{equation}
  \phi_{ab}=\frac{\langle ab\rangle}{[ab]}\quad(a\neq b)\,,\qquad\phi_{aa}=-\sum_{l\neq a}\frac{ [al]\langle lm\rangle\langle ls\rangle}{\langle al\rangle\langle am\rangle\langle as\rangle}\,.
\Label{eq:HodgesMatrix}
\end{equation}
Plugging \eqref{eq:Identities} into \eqref{eq:YMgravity} and then the integrated CHY formula  \eqref{eq:CHY}, we immediately get\footnote{Normalization factors are neglected.}
\begin{align}
A^{\text{YM}}(\cdots,x^-,\cdots,y^-,\cdots)\propto\frac{\langle xy\rangle^{4}} {\langle 12\rangle\cdots\langle n1\rangle}\,,&&A^{\text{GR}}(\cdots,x^-,\cdots,y^-,\cdots)\propto{\langle xy\rangle^{8}}\bar{M}(12\ldots n)
\end{align}
which are respectively the well known Parke-Taylor formula and the Hodges formula.

We prove that the other special solution $\{\overline{\sigma}_a\}$ does not contribute to MHV amplitudes~\cite{Du:2016blz}. In our following works, we prove that other solutions do not contribute to MHV amplitudes~\cite{Du:2016wkt,Du:2016fwe}. Thus the correspondence between the special solution $\{\sigma_a\}$ and the MHV amplitudes in Yang-Mills theory and gravity has been proved.

\section{Direct evaluation of single-trace MHV amplitudes in Einstein-Yang-Mills by CHY formula}
In this section, we review our discussion of the color-ordered  single-trace MHV amplitudes at tree level in EYM theory.
The color-ordered EYM amplitude is characterized by the number of gravitons $s$ and gluons $r$, with $s+r=n$.
At tree level, a single-trace color-ordered amplitude depends on permutations of external gluons (as the color-ordered pure Yang-Mills amplitudes), but not that of external gravitons (as the pure gravity amplitudes).
We use $\mathsf{h}$ and $\mathsf{g}$ to denote the sets of gravitons and gluons respectively. The set of all external particles thus is given by $\mathsf{p}=\mathsf{h}\cup\mathsf{g}=\{1,2,\ldots,n\}$. We use the following convention:
\begin{align*}
	&\mathsf{h}=\{1,2,\ldots,s\}\equiv\{h_1,h_2,\ldots,h_s\}\,,& &\mathsf{g}=\{s+1,s+2,\ldots,s+r\}\equiv\{g_1,g_2,\ldots,g_r\}\,.
\end{align*}
We will also use the sets of $+$ and $-$ helicity gravitons $\mathsf{h}_{\pm}$, gluons $\mathsf{g}_{\pm}$ and $\mathsf{p}_{\pm}=\mathsf{h}_{\pm}\cup\mathsf{g}_{\pm}$. The orders of these sets are denoted as $n=|\mathsf{p}|$, $s=|\mathsf{h}|$ and $r=|\mathsf{g}|$.

The CHY integrand for single-trace tree amplitude in EYM is given by \cite{CHY-EYM}
\bea
{\cal I}^{\text{EYM}}_{s,r}(h_{1},\ldots,h_{s},g_{{1}},\ldots,g_{{r}})=\frac{\text{Pf}\,(\Psi_{\mathsf{h}})\text{Pf}\,'(\Psi)}{\omega_{g_{1}g_{2}}\omega_{g_{2}g_{3}}\ldots\omega_{g_{r}g_{1}}},\Label{eq:EYM}
\eea
where $\Phi$ and $\Psi$ are already given by \eqref{eq:Phi} and (\ref{eq:Psi}) respectively, the same as in Yang-Mills case. The $\Psi_{\mathsf{h}}$ is an $2s\times 2s$ matrix and
given by:
\begin{equation}
	\Psi_{\mathsf{h}}(\{k,\W{\epsilon},\omega\})=\left(\begin{array}{cc}
		A_{\mathsf{h}} & -C_{\mathsf{h}}^{T} \\
		C_{\mathsf{h}} & B_{\mathsf{h}} \\
		\end{array}\right)\,.
\Label{eq:Psih}
\end{equation}
Here $A_{\mathsf{h}}$, $B_{\mathsf{h}}$ and $C_{\mathsf{h}}$ are $s\times s$ dimensional diagonal submatrices of $A$, $B$, and $C$ with indices in the graviton set $\mathsf{h}$. When we consider MHV amplitudes with two negative helicity particles, (as shown in the previous section) all other solutions make the reduced Pfaffian $\text{Pf}\,'(\Psi)$ in \eqref{eq:EYM} vanish. Thus we only need to substitute the special solution $\{\sigma_a\}$ into \eqref{eq:EYM}. According to which two particles take negative helicity, we have three different cases: $(g^{-}g^{-})$, $(h^{-}g^{-})$ and $(h^{-}h^{-})$.
\paragraph{$(g^{-}g^{-})$ amplitudes} 
We choose the polarization $\W{\epsilon}$ to be:
$\W{\epsilon}^{\mu}_{a}(+)=\langle q|\gamma^{\mu}|a]/(\sqrt{2}\langle qa\rangle)$ with $a\in\mathsf{h}$.
With this particular choice, all the reference vectors in $\W{\epsilon}$  are the same and all entries in the $ B_{\mathsf{h}}$ block of \eqref{eq:Psih} should vanish. The Pffafian $\text{Pf}\,(\Psi_{\mathsf{h}})$ is then given by
	$\text{Pf}\,(\Psi_{\mathsf{h}})=(-1)^{s(s+1)/2}\det({C}_{\mathsf{h}})\,$.
After plugging the special solution $\{\sigma_a\}$ into the above Pfaffian as well as the Parke-Taylor like factor, we arrive
\begin{equation} \frac{\text{Pf}\,[\Psi_{\mathsf{h}}(\sigma)]}{\sigma_{g_{1}g_{2}}\sigma_{g_{2}g_{3}}\ldots\sigma_{g_{r}g_{1}}}
=(-1)^{s(s-1)/2}(\sqrt{2}\,)^{s}F^{{n}}\left(P_{\xi}\right)^{2}\frac{\det(\phi_{\mathsf{h}})}{\langle g_{1}g_{2}\rangle\langle g_{2}g_{3}\rangle\ldots\langle g_{r}g_{1}\rangle}\,,
\Label{eq:Pfcf1}
\end{equation}
in which $\phi_{\mathsf{h}}$ is the $s\times s$ diagonal submatrix of the Hodges matrix \eqref{eq:HodgesMatrix} with all gluons rows and columns removed. Putting the first equation in \eqref{eq:Identities} and \eqref{eq:Pfcf1} together, we finally get the following expression of $(g^{-}g^{-})$ amplitudes
\begin{align}
	A_{s,r}^{\text{EYM}}(h_{1}^{+}\cdots h_{s}^{+};g_{{1}}^{+}\cdots g_{{i}}^{-}\cdots g_{{j}}^{-}\cdots g_{{r}}^{+})\propto\frac{\langle g_{i}g_{j}\rangle^{4}}{\langle g_{1}g_{2}\rangle\langle g_{2}g_{3}\rangle\ldots\langle g_{r}g_{1}\rangle}\det(\phi_{\mathsf{h}})\,.
\Label{eq:Mgg}
\end{align}

\paragraph{$(h^{-}g^{-})$ amplitudes} Suppose $h_i$ is the negative helicity graviton, we choose gauge as
\begin{align}
	&\W{\epsilon}^{\mu}_{i}(-)=\frac{\langle h_i|\gamma^{\mu}|q]}{\sqrt{2}[qh_i]}\,,& &\W{\epsilon}^{\mu}_{a}(+)=\frac{\langle h_i|\gamma^{\mu}|a]}{\sqrt{2}\langle h_ia\rangle}\,,& &(a\in\mathsf{h}_{+})\,,
\end{align}
which again makes all the entries of $ B_{\mathsf{h}}$ block in \eqref{eq:Psih} vanish. In addition, entries in the $i$-th column of $C_{\mathsf{h}}$, except for the diagonal element $C_{ii}$, are also zero. By substituting the special solution $\{\sigma_a\}$ into the Pfaffian $\text{Pf}\,(\Psi_{\mathsf{h}})$ and the Parke-Taylor like factor, we get the final expression of this amplitude
\begin{align}
	A_{s,r}^{\text{EYM}}(h_{1}^{+}\cdots h_{i}^{-}\cdots h_{s}^{+};g_{{1}}^{+}\cdots g_{{j}}^{-}\cdots g_{{r}}^{+})\propto\frac{\langle h_i g_{j}\rangle^{4}}{\langle g_{1}g_{2}\rangle\langle g_{2}g_{3}\rangle\ldots\langle g_{r}g_{1}\rangle}\det[(\phi_{\mathsf{h}})_{i}^{i}]\,.
\Label{eq:Mhg}
\end{align}
\paragraph{$(h^{-}h^{-})$ amplitudes} with two negative helicity gravitons $h_i$ and $h_j$, we choose gauge as
\begin{align}
	&\W{\epsilon}^{\mu}_{a}(-)=\frac{\langle a|\gamma^{\mu}|q]}{\sqrt{2}[qa]}\,,& &(a\in\mathsf{h}_{-}) &\W{\epsilon}^{\mu}_{a}(+)=\frac{\langle p|\gamma^{\mu}|a]}{\sqrt{2}\langle pa\rangle}\,,& &(a\in\mathsf{h}_{+})\,.
\Label{eq:gaugechoice}
\end{align}
By plugging the special solution and applying elementary transformations, we prove that the Pfaffian $\text{Pf}\,(\Psi_{\mathsf{h}})$ has to vanish. This result can be generalized: \emph{If gluons have the same helicity, we always have} $\text{Pf}\,(\Psi_{\mathsf{h}})=0$. This agrees with the conjectured result in~\cite{SBDW}.


%

\paragraph{Summary and comments}
Let us summarize the results for tree-level single-trace MHV amplitudes in EYM: The  $(h^{-}h^{-})$ amplitudes vanish while the $(g_i^{-}g_j^{-})$ and $(h_i^{-}g_j^{-})$ amplitudes can be expressed by the following new compact formula
\begin{equation}
	A_{s,r}^{\text{EYM}}(h_{1}^{+}\cdots i^{-}\cdots g_{j}^{-}\cdots g_{r}^{+})\propto\frac{\langle ig_j\rangle^{4}}{\langle g_{1}g_{2}\rangle\langle g_{2}g_{3}\rangle\ldots\langle g_{r}g_{1}\rangle}S(\mathsf{h}_{+}),\quad S(\mathsf{h}_{+})=(-1)^{|\mathsf{h}_+|}\det(\phi_{\mathsf{h}_{+}})\,.
\Label{eq:MHVGeneral}
\end{equation}
Another formula (SBDW) for tree-level single-trace EYM MHV amplitudes is proposed in~\cite{SBDW}:
%
\begin{align}
  S(\mathsf{h}_{+})=\left(\prod_{m\in\mathsf{h}_{+}}\frac{\partial}{\partial a_{m}}\right)\exp\left[\sum_{n_{1}\in\mathsf{h}_{+}}a_{n_{1}}\sum_{l\in\overline{\mathsf{h}_{+}}}\psi_{ln_{1}}\exp\left[\sum_{\substack{n_{2}\in\mathsf{h}_{+}\\n_{2}\neq n_{1}}}a_{n_{2}}\psi_{n_1n_2}\exp\left(\cdots\right)\right]\right]_{a_{m}=0}\,,
\Label{eq:Sh+}
\end{align}
where $\psi_{ab}=\phi_{ab}\langle b\xi\rangle\langle b\eta\rangle/(\langle a\xi\rangle\langle a\eta\rangle)$. Here $\overline{\mathsf{h}_{+}}$ is the complement of $\mathsf{h}_{+}$ in $\mathsf{p}=\{1,2,\ldots,n\}$.
In fact the two formulas \eqref{eq:MHVGeneral} and \eqref{eq:Sh+}  are equivalent with each other. This equivalence are proved by developing the following new \emph{spanning forest theorem} \cite{Du:2016wkt}
\bea
 S(\mathsf{h}_{+})=\sum_{F\in\mathcal{F}_{\overline{\mathsf{h}_{+}}}(K_n)}\left(\prod_{v_av_b\in E(F)}\psi_{ab}\right)\,.
\eea
Here we summed over all possible forests $F\in\mathcal{F}_{\overline{\mathsf{h}_{+}}}(K_n)$. Each $K_n$ is a weighed complete graph with vertices $\{v_{1},\ldots,v_{n}\}$ and weight $\psi_{ab}$ is assigned to the edge $v_av_b$.  $\mathcal{F}_{\overline{\mathsf{h}_{+}}}(K_n)$ means that we only consider those forests $K_n$ rooted on the vertices assigned by $\overline{\mathsf{h}_{+}}$. All diagrams are assumed to be directed away from the roots. On the other hand, as proved in \cite{Feng:2012sy},  $(-1)^{|\mathsf{h}_+|}\det(\phi_{\mathsf{h}_{+}})$ produces the same graphical rule. Thus the two formulas \eqref{eq:MHVGeneral} and \eqref{eq:Sh+} precisely match with each other.

\section{Characterization of solutions of scattering equations}

To study amplitudes beyond MHV by direct evaluation of the integrated CHY formula, we need more information about other solutions of scattering equations. Inspired by the direct evaluation of CHY formula, we propose a rank characterization of solutions of scattering equations \cite{Du:2016fwe}\footnote{Some discussions are overlap with other independent works \cite{He:2016iqi}. }. Particularly, we define the following discriminant matrices $\mathfrak{C}_{\pm}$,
 \begin{align}
&(\mathfrak{C}_{-})_{ab}=\left\{\begin{array}{>{\displaystyle}l @{\hspace{1em}} >{\displaystyle}l}
\frac{\langle ab\rangle}{\omega_{ab}} & a\neq b\\
-\sum_{{b\neq a}}\frac{\langle ab\rangle[bq]}{\omega_{ab}[aq]} & a=b \\
\end{array}\right.\,,
& &(\mathfrak{C}_{+})_{ab}=\left\{\begin{array}{>{\displaystyle}l @{\hspace{1em}} >{\displaystyle}l}
\frac{[ab]}{\omega_{ab}} & a\neq b\\
-\sum_{{b\neq a}}\frac{[ab]\langle bp\rangle}{\omega_{ab}\langle ap\rangle } & a=b \\
\end{array}\right.\,,
\Label{eq:Cpm}
\end{align}
and prove that only those solutions in the subset labeled by $\rank(\mathfrak{C}_{-})=k+1$ support the $\text{N}^{k}\text{MHV}$ amplitudes. Solutions characterized by this method
are shown to have an Eulerian number pattern \cite{Du:2016fwe}, which was also observed and understood in other ways \cite{EulerianNumber}. A consequent result of this characterization is that if the gluons in the tree-level single-trace EYM amplitudes have the same helicity, the amplitudes must vanish identically \cite{Du:2016fwe}.

\section{Conclusions}
In this talk, we reviewed direct evaluation of the integrated CHY formula for tree amplitudes in Yang-Mills theory, gravity and Einstein-Yang-Mills theory. By comparing the results with Parke-Taylor and Hodges formulas, we explicitly prove that only the MHV solution support the MHV amplitudes in Yang-Mills theory and gravity. We also proposed a new compact formula for $(g^{-}g^{-})$ and $(h^{-}g^{-})$ tree-level MHV amplitudes in the single-trace sector of Einstein-Yang-Mills theory. By the use of graphical rule, we proved the equivalence between this new formula and the SBDW formula. The $(h^{-}h^{-})$ amplitudes are proved to be zero. All these direct evaluations support the correspondence between MHV solution and MHV amplitudes. Inspired by direct evaluation of CHY formula, we introduced a characterization method for general solutions of scattering equations.

\end{document}